\documentstyle[emulateapj,apjfonts,flushrt]{article}
\makeatletter

\newlength{\bibhang}
\let\@internalcite\cite
\def\cite{\@ifstar{\citeyear}{\citefull}}
\def\cite{\let\@citeleft(\let\@citeright)%
    \@ifstar{\citeyear}{\citefull}}
\def\citenp{\let\@citeleft\relax\let\@citeright\relax
    \@ifstar{\citeyear}{\citefull}}
\def\citefull{\def\astroncite##1##2{##1~##2}\@internalcite}
\def\citeyear{\def\astroncite##1##2{##2}\@internalcite}

\def\@citex[#1]#2{\if@filesw\immediate\write\@auxout{\string\citation{#2}}\fi
  \def\@citea{}\@cite{\@for\@citeb:=#2\do
    {\@citea\def\@citea{; }\@ifundefined
       {b@\@citeb}{{\bf ?}\@warning
       {Citation `\@citeb' on page \thepage \space undefined}}%
{\csname b@\@citeb\endcsname}}}{#1}}

\def\@cite#1#2{\@citeleft#1\if@tempswa , #2\fi\@citeright}
\def\@biblabel#1{}

\makeatother

\newcommand{\PSbox}[3]{\mbox{\rule{0in}{#3}\includegraphics{#1}\hspace{#2}}}
\newcommand{\FigNum}[1]{\unitlength 1pt \begin{picture}(55,10)(-400,35) 
                        \put(0,0){Figure #1}
                        \end{picture}}
\newcommand{\perval}[2]{{#1\mbox{$^{#2}$}}}
 
\newcommand{\persec}{\perval{sec}{-1}\/}
\newcommand{\percm}{\mbox{$\cm^{-2}$}}

\newcommand{\ppm}{\mbox{$\pm$}}
\newcommand{\cgsflux}{\erg~\percm~\persec}
\newcommand{\cgslum}{\erg~\persec}

\newcommand{\approxlt}{\mbox{$\lesssim$}}

\def\etal{{et~al.}}

\newcommand{\ud}[2]{\mbox{$^{+ #1}_{- #2}$}}

\newcommand{\ee}[1]{\mbox{$10^{#1}$}}
\newcommand{\tee}[1]{\mbox{$\times 10^{#1}$}}

\newcommand{\cm}{\mbox{$\rm\,cm$}}
\newcommand{\km}{\mbox{$\rm\,km$}}

\newcommand{\erg}{\mbox{$\rm\,erg$}\/}

\newcommand{\chandra}{{\em Chandra\/}}
\newcommand{\rosat}{{\em ROSAT\/}}

\newcommand{\lp}{{LP~944-20}}
\newcommand{\usno}{{USNO-A2}}
\newcommand{\hal}{{H$\alpha$}}

\newcommand\lxlbol{$L_{X}$/$L_{\rm bol}$}  
\newcommand\lx{$L_{X}$}  

\received{April 28, 2000}
\revised{May 17, 2000}
\slugcomment{\it ApJ Letters, Submitted April 28, 2000; Revised May 17, 2000}
\begin{document}

\title{\chandra \ Detection of an X-ray Flare from the Brown Dwarf \lp}
\author{Robert E. Rutledge}
\affil{Division of Physics, Mathematics and Astronomy}
\affil{MS 220-47, California Institute of Technology, Pasadena, CA
91107\\ {\em rutledge@srl.caltech.edu} }
\smallskip
\author{Gibor Basri}
\affil{Department of Astronomy, University of California at Berkeley}
\affil{Berkeley, CA 94720-3411\\ {\em basri@soleil.berkeley.edu}}

\author{Eduardo L. Mart\'\i n}
\affil{Division of Geological and Planetary Sciences, California
Institute of Technology}
\affil{MS 150-21,
Pasadena, CA 91125\\ {\em ege@gps.caltech.edu}}

\centerline{\sc and}
\author{Lars Bildsten}
\affil{Institute for Theoretical Physics and Department of Physics}
\affil{Kohn Hall, University of California, Santa Barbara, CA 93106\\ {\em bildsten@itp.ucsb.edu}}

\begin{abstract}

   We have detected a bright X-ray flare from the nearby ($d=$5.0 pc)
brown dwarf \lp\ with the \chandra/ACIS-S. This is an old (500 Myr),
rapidly rotating, lithium-bearing M9 object, with a bolometric
luminosity of $\approx 6$\tee{29} \cgslum. It was only detected by
{\it Chandra} during an X-ray flare of duration 1-2 hours near the end
of a 12.1 hour observation.  The peak X-ray luminosity was
1.2\ud{0.5}{0.3}\tee{26} \cgslum\ in the brightest $\approx 550$
seconds, corresponding to $L_X/L_{\rm bol}\approx 2\times 10^{-4}$.  A
total of 2\tee{29} ergs was released during the 43,773 sec
observation, giving a time-averaged $L_X/L_{\rm bol}\approx 7\times
10^{-6}$.  \lp\ was not detected before the flare, with a 3$\sigma$
upper limit on the emission at $L_X/L_{\rm bol}< 2\times 10^{-6}$
($L_X<$1\tee{24} \cgslum).  This is faint for a rapidly rotating
late-type star, and establishes a record lower limit to the quiescent
flux about an order of magnitude below the flux limit (and a factor of
5 below the \lxlbol\ limit) placed on quiescent X-ray emission from
the M8 dwarf VB 10. The inferred flaring duty cycle is comparable to
that measured via variable H$\alpha$ emission for other late M-type,
fully convective stars.

\end{abstract}


\section{Introduction}

\lp\ (=BRI 0337$-$3535) is an isolated, non-accreting, brown dwarf
identified through its Li abundance and low luminosity
\cite{tinney98discovery}.  Its parallactic distance and bolometric
luminosity are 5.0\ppm0.1 pc and 6\tee{29}\cgslum
\cite{tinney96}. Tinney ~\cite*{tinney98discovery} infers an age of
about $\tau=$500 Myr, which implies that it is a fully collapsed
object (see \citenp{feigelson99} for a recent review of protostellar
evolution).  This requires that any observed coronal activity not be
due to accretion, as may power younger brown dwarfs ($\tau<$10 Myr) in
analogy with T-Tauri stars.  We report here on a bright 1-2 hour X-ray
flare detected with \chandra \ during a 12 hour observation, and a
strong upper limit on persistent X-ray emission. We attribute the
energy release in the flare to transient magnetic activity on this
fully convective, rapidly rotating star.

Stars with masses $M>0.3 M_\odot$ have an outer convective zone and
an interior radiative region that need not be rotating at the same
rate. A poloidal magnetic field in the convective layers will be
stretched and amplified into strong toroidal fields when it is dragged
by convective overshoot (see \citenp{weiss94}) into the radial shear
in rotation that resides at the boundary (in and near the so-called
"tachocline"; \citenp{spiegel92}).  In these cases, rapid rotation is
associated with enhanced coronal activity. The activity level
correlates with the Rossby number -- the ratio of the stellar rotation
period to the convective turnover time \cite{Noyes84}. For Rossby
numbers between 10 and 0.1, coronal activity (as measured by \lxlbol)
increases with decreasing Rossby number (that is, with more rapid
rotation), ``saturating'' at about \lxlbol $\sim 10^{-3}$ for Rossby
numbers of 0.1 to 0.01 \cite{randich98}.

 For less massive stars and young brown dwarfs, the energy is
transported throughout the star by convection; no radiative core is
present. For this reason, it has been supposed that the activity and
its dependence on rotation might change near the spectral type where
the radiative layer disappears (about M5.5; see \citenp{giampapa96}
for a physical overview).  However, a search for this effect found no
evidence for a change in the saturated value of \lxlbol $\sim
10^{-3}$, down to spectral types as late as M7, well into the region
of fully convective stars \cite{fleming93}.  This implied no dramatic
change in the rotation-activity dependence in fully convective stars.

The first hint that activity might be decreasing in the very late M
dwarfs was observed from the rapidly rotating ($v \sin i = 40$ \km\
\persec) M9.5 star BRI 0021-0214, for which a strong upper-limit on
\hal\ emission indicated a substantially lower persistent coronal
activity than expected from a rapidly rotating, fully convective star
\cite{basri95}.  On the other hand, Reid \etal\ \cite*{reid99dec}
observed a strong \hal\ flare from BRI 0021-0214; this indicated that
activity was indeed present, and that the star has outbursts no more
than 7\% of the time. Liebert \etal\ ~\cite*{liebert99} reported a
bright \hal\ flare from 2MASSW J0149090+295613, an M9.5 V star which
is otherwise quiescent. Intensity variability from brown dwarfs has
been searched for, with limited success.  Bailer-Jones \& Mundt
\cite*{bailerjones99} found no infra-red variability in three Pleiades
brown dwarfs, with a limit of $\delta I<$0.05 mag on timescales
between 25 min-27 hrs.  A search for ``weather'' in two brown dwarfs
(\lp\ and DENIS-P J1228-1547) produced a claim of evidence of
variability in one, at the 2.3$\sigma$ level \cite{tinney99}.

Previous X-ray observations of brown dwarfs and brown dwarf candidates
have produced detections of persistent emission from several
\cite{neuhauser98,neuhauser99a}, but these are all young objects
(\approxlt 10 My), still in the process of proto-stellar collapse, and
so are actively accreting; neither are they as cool or faint as older
brown dwarfs such as \lp.  The activity observed from these collapsing
brown dwarfs is analogous to that in the also-young T-Tauri stars --
powered by accretion and collapse -- rather than that in fully
collapsed, older M dwarfs.  There are only a few observations of X-ray
emission in older late M dwarfs and brown dwarfs. Giampapa
\etal~\cite*{giampapa96} reported on a ROSAT detection from VB 8
(spectral type M7) with a time-averaged \lxlbol =1.6\tee{-3}. Fleming
\etal~\cite*{fleming00} detected the M8 star VB 10 during a flare at
\lxlbol =5\tee{-4}. No quiescent X-ray emission was detected, limiting
it to \lxlbol $<$\ee{-5}. In the only previous observations of \lp ,
an upper limit of \lxlbol $<$7\tee{-5} was found
\cite{neuhauser99a}. Hence, as with \hal\ observations, the X-ray
behavior of late M stars tends toward flaring activity and an absence
of persistent activity at the X-ray detection limits of present
instrumentation.  Our work confirms this tendency.

We report here on our \chandra \ observation of an X-ray flare from
the brown dwarf \lp.  In \S~\ref{sec:obs}, we describe the
observation, analysis and detected flux level. In \S 3, we interpret
the light curve of the flare.  We conclude in \S~\ref{sec:con} with a
brief summary of our results and a comparison to other work. 

\section{Chandra Observation and Analysis}
\label{sec:obs}

The observation (\chandra\ sequence number 200049) occurred on 15 Dec
1999 00:05:50-13:03:05 UTC, for 43,773 seconds. \lp\ was targeted at
the nominal aimpoint for the ACIS-S3 chip (backside illuminated) in a
faint imaging mode, with 3.2 sec time resolution.  Analysis of the
ACIS-S3 chip countrate during the observation showed no evidence of
the background flares that sometimes appear (ACIS background
calibration memo
\footnote{http://asc.harvard.edu/cal/Links/Acis/acis/Cal\_prods/bkgrnd/11\_18/bg181199.html}),
on timescales longer than 50 sec, greater than a factor of $\sim$few
(background variability is discussed in more detail in
Sec.~\ref{sec:var}). The standard analysis data products from \chandra\
found 117 X-ray objects in the field of view.

  We now describe how we have used the observations to measure the 
X-ray emission from \lp. 

\subsection{Astrometry and Source Identification}

The X-ray source closest to the ACIS-S detector aimpoint was offset
from the detector aimpoint by $\delta$RA=$-2.2$\arcsec and
$\delta$dec=$-$8.3\arcsec\ according to the standard product
astrometry.  This offset is consistent with known systematic
uncertainties in the \chandra\ pre-processing analysis astrometry, of
the version which produced the astrometry for this observation.  We
performed astrometry using the absolute positions of objects in \usno.
We used the \chandra\ Interactive Analysis of Observations (CIAO) V1.0
software tool {\em celldetect} to find X-ray point source relative
positions using counts detected in PHA channels 10-400 (correspondig
roughly to 0.1-4.0 keV). We found 39 X-ray sources within 10\arcmin\
of the aimpoint, with relative astrometry accurate to between
0.03-0.18\arcsec.  We extracted all optical source positions (from the
USNO-A2 catalog) within 15\arcsec\ of the X-ray source positions. We
did the same for 20 background fields for each X-ray source (a total
of 780 fields), offset by increments of 15\arcsec\ from each X-ray
source position, to find the USNO-A2 source density in the region of
$\rho_{\rm USNO-A2}=(5.2\ppm0.3)\tee{-4}$ \perval{arcsec}{-2}.  Of the
39 \chandra\ X-ray sources, 12 had a \usno\ object within 15\arcsec.
Of these 12, seven were within 1.0\arcsec\ of the offset
$\delta$RA=$-2.2$\arcsec and $\delta$dec=$-$8.3\arcsec.  The
probability of n=7 (of m=12) optical point sources being found within
$r=1.0$\arcsec\ of a previously selected position is $P(<r)=
\frac{m!}{(m-n)!n!}(1 - \exp(-\pi \rho_{\rm USNO-A2}
r^2))^n=$2\tee{-17}.  Thus, the clustered \usno\ sources are -- taken
together -- likely to be the optical counterparts for their
corresponding X-ray sources, useful for astrometry.

To calculate the systematic shift in RA and Dec, we adopted the
relative positional uncertainties for the X-ray sources found by {\em
celldetect}, and adopted absolute positional uncertainties of
0.25\arcsec\ for USNO-A2 sources.  

The astrometric correction is ($\delta$RA=$+2.2$\arcsec,
$\delta$dec=8.4\arcsec), with an uncertainty of \ppm0.1 \arcsec.  The
astrometricly corrected X-ray position of the aimpoint source is then
RA=03h39m35.16s, dec=$-$35d25h44.0s \ppm 0.1\arcsec\ (1$\sigma$;
J2000, epoch 1999.95).  The positional difference between the aimpoint
source and the optical position of \lp\ at this epoch \cite{tinney96}
is $\delta$RA=$-0.67\pm0.23$\arcsec\ and
$\delta$dec=0.15\ppm0.23\arcsec, which is consistent at the 3$\sigma$
level.

The likelihood  of a chance alignment of  a serendipitous X-ray source
within $r$=1\arcsec\ of  an arbitrary position is  $P(<r)=0.0002$, for
$\rho_{\rm X-ray}=$7.2\tee{-5}  \perval{arcsec}{-2}, found from  the 18
sources on the ACIS-S3 chip.  We therefore  identify this X-ray source
with the brown dwarf \lp\ with 99.98\%  confidence on the basis of the
positional coincidence.

\subsection{The Flare from \lp} 
\label{sec:var} 

First, we estimate the background countrate. We obtained background
counts from an annulus centered on the source, with inner- and
outer-radius of 4 and 140 pixels.  We excluded data within 10 pixels
of two X-ray sources localized by {\em celldetect} to be within this
annulus.  The total average background countrate was
(3.99\ppm0.04)\tee{-6} counts \persec \perval{pixel}{-1}; in the
limited PHA channel range of 10-400 (nominally 0.1-4.0 keV), the
background countrate was (1.0\ppm0.02)\tee{-6} counts \persec
\perval{pixel}{-1}.  For nearly all the analyses we present herein,
this level of background is negligible.

We extracted counts from a circle about \lp\ 2.0 pixels in radius
(0.98\arcsec-- the 90\% enclosed energy radius at 1.5 keV, on axis).
The total number of expected background counts in the source region is
2.27\ppm0.02 counts (0.1-10.0 keV), and 0.57\ppm0.01 counts (0.1-4.0
keV).  We produced a (0.1-4.0 keV) light curve of 552 sec time
resolution (Fig.~\ref{fig:sub4varicomp}).  Of the 19 detected counts
in the source region, 15 were detected during a 2760 sec period
beginning at 1999 Dec 15 09:41:25 (UTC), which we arbitrarily define
as the ``flare period''.  Of these, 7 were detected in a single 552
sec bin.  In 80 such time bins, with an average of 19/80=0.238
counts/bin, the probability of randomly finding 7 counts in one of 80
bins (assuming a constant countrate) is $\approx 7\tee{-9}$.

The ACIS-S-BI chips suffer from short-term increases in background
countrate; these can increase the chip background countrate by up to
factors of 100 across the entire chip, and in all PHA channels on
timescales of seconds to minutes.  The ratio of the countrates during
and outside the flare period in the source region is 18-200 ($3\sigma$
range, assuming Poisson statistics), while in the background region it
is 0.85-1.07 (3$\sigma$, assuming Gaussian statistics).  The flare
cannot be from a variation in the background and must be a change in
the X-ray emission of \lp.

\subsection{Spectral Analysis}

There were 15 counts (0.1-10 keV) in the 2.0 pixel (=0.98\arcsec)
radius about \lp\ during the 2760.0 seconds of the flare.  We expect
0.14 background counts, which we neglect for this spectral analysis.

We performed a spectral analysis, binning the data into two bins which
contained 7 and 8 counts, with a third bin of PHA channels 95-1024
which contained no counts.  We fit the resulting spectrum with an
assumed Raymond-Smith plasma model, implemented in XSPEC \cite{xspec};
the best fit ($\chi^2/\nu$=0.003, for $\nu$=1 degree of freedom) found
$kT$=0.26\ud{0.1}{0.07} \ {\rm keV}(90\% confidence), and a
time-averaged flux during the 2760 seconds of (1.5\ppm0.4)\tee{-14}
\cgsflux (\lx=[4.5\ppm1.2]\tee{25} \cgslum; 0.1-10 keV), which
corresponds to a flux conversion factor of 1 count = 2.8\tee{-12}
\erg\percm \ (0.1-4.0 keV) for this spectrum.  This reduced chi-square
value is small because the high energy bin is essentially unimportant,
as there are, in the best fit model, $\sim$0 counts in this bin, and
thus the model can be thought of as completely determined, with only
two bins and two unknowns, although the high energy bin does provide a
constraint on flatter spectra, and thus on $kT$ and flux
normalization.

The uncertainty in this conversion factor depends both on the
uncertainty in the Raymond-Smith spectral parameters, and on the
uncertainty in the intrinsic spectrum.  The best fit power-law
spectrum ($\alpha_{photon}$=2.6), however, is acceptable only at 6\%
confidence. The best-fit black body spectrum (kT=0.17 keV) gives a
flux 10\% lower than the best-fit Raymond Smith spectrum.  We estimate
this as the level of spectral uncertainty in the flux conversion
factor. 

In Table~\ref{tab:flux}, we list the \lxlbol\ values for the full
observational period, pre-flare period, the flare period, and the peak
550 sec bin of the flare period.  The uncertainties in these values
are Poisson (counting statistics), plus spectral uncertainty
($\sim$10\%).  The $L_X$ were obtained from the detected countrates,
corrected for background and the Enclosed Energy fraction appropriate
for our source region ($0.90$), and using the above flux conversion
factor.

The total source fluence (including the flare) during this observation
is 5.7\tee{-11} ergs \percm, corresponding to a released energy of
1.7\tee{29} ergs, and a time-averaged luminosity of
$L_X$=4\tee{24}\cgslum.  The time-averaged \lxlbol\ is then 7\tee{-6}.
The peak luminosity during the flare was 1.2\ud{0.5}{0.3}\tee{26}
\cgslum (including only the counting statistical uncertainty, not the
10\% systematic uncertainty in the counts-energy conversion factor),
corresponding to $L_X/L_{\rm bol}\approx 2\times 10^{-4}$.

Before the flaring period, there is 1 count (0.1-4 keV) in 34535 sec,
where 0.45 background counts are expected (probability p=0.36 -- consistent with
background).  The 3$\sigma$ upper-limit on the pre-flare flux is
$<$4.5\tee{-16} \cgsflux\ ($L_X<$1\tee{24} \cgslum;
\lxlbol$<$2\tee{-6}).  This compares to the VB~10 quiescent coronal
X-ray limits from Fleming \etal\ \cite*{fleming00} of
$L_X<$1.7\tee{25}\cgslum, which is an order of magnitude greater, and
\lxlbol$<$1.0\tee{-5}, which is a factor of $\times$ 5 greater than
we find here for \lp.

After the flare period, there are 3 counts in 6478 sec (0.1-4 keV)
where 0.084 are expected (p=0.0001).  Thus, there are significant
counts after the flare period, which indicates that the flare event
continues beyond the flaring period we defined.

\section{Constraints on the Form of the Lightcurve}

Due to the low number of counts, it is difficult to measure the
lightcurve parameters of this flare in a model-independent way.  We
therefore imposed a particular model and measured the resulting
parameters.  We imposed a model of an instantaneous rise/exponential
decay flare, set above a constant background countrate, to compare
with the 0.1-4 keV energy band data (19 counts).  We assumed the
background countrate from our background region.  We asserted that the
instantaneous rise takes place at the time of the first detected
photon in the vicinity of the flare (32969 seconds into the
observation).  The resulting lightcurve (background+flare) must
produce between 11-27 counts in total (90\% confidence region of the
total number of counts observed).  The distribution of counts in time
should match that observed over the full observation period, using a
Kolmogorov-Smirnov test \cite{press}, that the data and the model be
consistent at the 10\% level.  We tested a grid of peak countrate
($I_p$) between \ee{-4} and 1 counts~\persec, in units of \ee{-5}
counts~\persec, and exponential decay timescale ($\tau$) between 1
and 2\tee{4} in units of 100 sec.  Based on this model and approach,
we find a 90\% confidence region for $\tau$=5400\ppm700 s, and
$I_p$=0.002-0.006 c \persec.  When we relax the constraint on start
time, permitting the flare to begin at any time between 30\tee{3} and
40\tee{3} sec after observation start in grid units of 100 sec, the
90\% confidence interval of $\tau$ is 500-7100 s, and of $I_p$ is
0.0018-0.054 counts~\persec.

This brown dwarf is known to be rapidly rotating, with a measured
$V\sin i\approx 28 \ {\rm km \ s^{-1}}$ \cite{tinney98spec}.  For the
reported bolometric luminosity of $L_{\rm Bol}\approx
(5.5\ppm0.4)\times 10^{29}$\cgslum\ \cite{tinney98discovery} and
effective temperature of 2500\ppm100 K \cite{basri00}, the inferred
radius is $R\approx (4.5\pm0.4)\times 10^{9} \ {\rm cm}$, which is
35\ppm10\% smaller than the $R=0.1R_\odot$ expected for an object of
this mass and age \cite{burrows97}. The expected radius could be
derived if the temperature were lowered to 2300K and the bolometric
luminosity increased to $8\times 10^{29}$\cgslum. Such changes are
within the observational uncertainties.  Presuming a radius of $R=0.1
R_\odot$, the rotation period is $4.4 \ {\rm hr} \sin i$.  One might
interpret the X-ray lightcurve as a localized flare that rotates out
of view for a period and then re-appears. The implied rotation rate of
$\approx$ 1 hour does not require an especially unlikely value of $\sin
i$.  However, such an interpretation is by no means required by the
data to explain the variations, as these are entirely consistent with
the uncertainty in the flare time evolution and Poisson counting
statistics.

\section{Discussion and Conclusions}
\label{sec:con}

 We have detected an X-ray source which we identify with the brown
dwarf \lp, with 99.98\% confidence, based on positional
coincidence. The detection was during a flare, with a peak X-ray
luminosity of $L_X=$1\ud{0.5}{0.3}\tee{26} \cgslum, or $L_X/L_{\rm
bol}\approx 2\times 10^{-4}$.  This flare's peak luminosity is a
factor of ten below that of an X-ray flare detected from the
comparably luminous M8 star VB~10 ($L_{\rm Bol}\approx 1.7 \times
10^{30} \ {\rm erg \ s^{-1}}$; \citenp{fleming00}). In VB~10, the peak
X-ray luminosity exceeded  $10^{-3} L_{\rm Bol}$. 

We have analysed the X-ray light curve with a simple sharp rise plus
exponential decay model; this indicates a decay time $5400\pm700 \
{\rm s}$ when we specify the flare start-time, and in the range of
500-7100 s when we do not specify the flare start time. This implies
\lp\ is flaring $\sim$10\% of the time, comparable to the amount of
\hal\ flaring among the coolest M dwarfs \cite{gizis00}, although the
uncertainty on our value is large.  The light-curve is consistent with
rotational modulation, but the counting statistics are too poor to
conclude that it is required, and discovery of such modulation must
wait for higher signal-to-noise data. A previous attempt to detect
this source in X-rays \cite{neuhauser99a} placed an upper-limit on the
time-averaged luminosity ($L_X< 4\times 10^{25} \ {\rm erg \ s^{-1}}$)
during a 66 ksec \rosat/PSPC observation and a 220 ksec \rosat/HRI
observation. These limits are above the time-averaged luminosity we
detect here ($L_X\approx 4\times 10^{24} \ {\rm erg \ s^{-1}}$), and
are therefore consistent with our results.  X-ray variability
was not discussed by these previous works; however, a cursory review
of 220~ksec of ROSAT/HRI data revealed no variability on timescales of
500 sec; the detection sensitivity (5 counts in a 500 sec bin, energy
conversion factor of 3.2\tee{-11} \cgsflux\ \perval{count}{-1}), is
about a factor of 10 above the present peak-flare detection.

The time-averaged luminosity ($L_X$=4\tee{24}\cgslum) is a factor of
$10^4$ below the previous detections of brown dwarfs and brown dwarf
candidates; the time-averaged ratio \lxlbol(=7\tee{-6}) is factor of
$10^2$ below this ratio \cite{neuhauser98,neuhauser99a}.  This might
be due to different energy production mechanisms; the previously
detected brown dwarfs are young compared with \lp\ ($\sim$1 My vs. 500
My) and are still forming, while \lp\ is isolated and undergoing slow
gravitational contraction.  The persistent X-ray emission of the very
young brown-dwarfs in open clusters is more analogous to the pre-main
sequence T-Tauri stars -- where convection is stronger, the
atmospheres are less neutral, and accretion may play a role -- while
the flaring X-ray emission of \lp\ is more analogous to the flaring
emission of late-type main-sequence stars, such as VB~8 and VB~10.

We estimate (order of magnitude) an eddy turnover time of $\sim$1 year
for \lp, yielding a Rossby number of $R_0 \sim$5\tee{-4}.  As
discussed earlier, this is below the Rossby number limit at which the
X-rays are observed to be "saturated" (at \lxlbol$\sim$\ee{-3}). That
saturation level is much greater than our quiescent X-ray flux upper
limit.  Either the "supersaturation" suggested by Randich has strongly
set in, or the connection between Rossby number and observed activity
is no longer relevant.  It is possible that the rapid rotation
suppresses persistent coronal activity, either by forcing the magnetic
field into a more organized form or by suppressing the turbulent
dynamo.  It is also possible that the neutrality of a cool photosphere
quenches coupling between atmospheric motions and the magnetic field,
which forces the field into dissipative configurations (as pointed out
by \citenp{fleming00}).  In any case, the detected flare requires that
a magnetic field be present on LP 944-20, and that at least
occasionally it is forced into a dissipative configuration high in the
atmosphere.

   Our results confirm the impression from previous studies of \hal\
that stellar activity is dying at the bottom of the main sequence, at
least in the form that it has in more massive late-type stars.  We
have pushed the limits on quiescent coronal emission levels to new
lows for fully convective objects. We have also helped confirm that
such objects do apparently still have magnetic fields and an ability
to flare about 10\% of the time.

\acknowledgements

The authors are grateful to the \chandra\ Observatory team for
producing this exquisite observatory. We are grateful to Andrew
Cumming for his critical reading of an early version of this paper,
and to the referee Tom Fleming, for comments which dramatically
improved the paper's readability.  We thank Daniel Holz for
encouraging us to consider the rotational modulation of the
lightcurve. This research was supported by NASA Grant No. GO0-1009X
and the National Science Foundation under Grant No.  PHY94-07194.
L. B. is a Cottrell Scholar of the Research Corporation.


\begin{thebibliography}{}

\bibitem[\protect\astroncite{Arnaud}{1996}]{xspec}
Arnaud, K.~A., 1996,
\newblock in G. Jacoby \& J. Barnes (eds.), {\em Astronomical Data Analysis
  Software and Systems V.}, Vol. 101, p.~17, ASP Conf. Series

\bibitem[\protect\astroncite{{Bailer-Jones} \& {Mundt}}{1999}]{bailerjones99}
{Bailer-Jones}, C. A.~L. \& {Mundt}, R., 1999,
\newblock {\em \aap} { 348}, 800

\bibitem[\protect\astroncite{{Basri} \& {Marcy}}{1995}]{basri95}
{Basri}, G. \& {Marcy}, G.~W., 1995,
\newblock {\em \aj} { 109}, 762

\bibitem[\protect\astroncite{{Basri} {\rm et~al.\/}}{2000}]{basri00}
{Basri}, G., {Mohanty}, S., {Allard}, F., {Hauschildt}, P.~H., {Delfosse}, X.,
  {Mart\'\i n}, E.~L., {Forveille}, T., \& {Bertrand}, G., 2000,
\newblock {\em \apj},
\newblock in press, astroph/0003033

\bibitem[\protect\astroncite{{Burrows} {\rm et~al.\/}}{1997}]{burrows97}
{Burrows}, A., {Marley}, M., {Hubbard}, W.~B., {Lunine}, J.~I., {Guillot}, T.,
  {Saumon}, D., {Freedman}, R., {Sudarsky}, D., \& {Sharp}, C., 1997,
\newblock {\em \apj} { 491}, 856

\bibitem[\protect\astroncite{{Feigelson} \& {Montmerle}}{1999}]{feigelson99}
{Feigelson}, E.~D. \& {Montmerle}, T., 1999,
\newblock {\em \araa} { 37}, 363

\bibitem[\protect\astroncite{{Fleming} {\rm et~al.\/}}{2000}]{fleming00}
{Fleming}, T.~A., {Giampapa}, M.~S., \& {Schmitt}, J. . H. M.~M., 2000,
\newblock {\em \apj} { 533}, 372

\bibitem[\protect\astroncite{{Fleming} {\rm et~al.\/}}{1993}]{fleming93}
{Fleming}, T.~A., {Giampapa}, M.~S., {Schmitt}, J. H. M.~M., \& {Bookbinder},
  J.~A., 1993,
\newblock {\em \apj} { 410}, 387

\bibitem[\protect\astroncite{{Giampapa} {\rm et~al.\/}}{1996}]{giampapa96}
{Giampapa}, M.~S., {Rosner}, R., {Kashyap}, V., {Fleming}, T.~A., {Schmitt}, J.
  H. M.~M., \& {Bookbinder}, J.~A., 1996,
\newblock {\em \apj} { 463}, 707

\bibitem[\protect\astroncite{Gizis {\rm et~al.\/}}{2000}]{gizis00}
Gizis, J.~E., Monet, D.~G., Reid, N.~I., Kirkpatrick, J.~D., Liebert, J., \&
  Williams, R.~J., 2000,
\newblock {\em \aj},
\newblock in press, astro-ph/0004361

\bibitem[\protect\astroncite{{Liebert} {\rm et~al.\/}}{1999}]{liebert99}
{Liebert}, J., {Kirkpatrick}, J.~D., {Reid}, I.~N., \& {Fisher}, M.~D., 1999,
\newblock {\em \apj} { 519}, 345

\bibitem[\protect\astroncite{{Neuh\"auser} {\rm et~al.\/}}{1999}]{neuhauser99a}
{Neuh\"auser}, R., {Brice\~no}, C., {Comer\'on}, F., {Hearty}, T., {Mart\'\i
  n}, E.~L., {Schmitt}, J. H. M.~M., {Stelzer}, B., {Supper}, R., {Voges}, W.,
  \& {Zinnecker}, H., 1999,
\newblock {\em \aap} { 343}, 883

\bibitem[\protect\astroncite{{Neuh\"auser} \& {Comer\'on}}{1998}]{neuhauser98}
{Neuh\"auser}, R. \& {Comer\'on}, F., 1998,
\newblock {\em Science} { 282}, 83

\bibitem[\protect\astroncite{{Noyes} {\rm et~al.\/}}{1984}]{Noyes84}
{Noyes}, R.~W., {Hartmann}, L.~W., {Baliunas}, S.~L., {Duncan}, D.~K., \&
  {Vaughan}, A.~H., 1984,
\newblock {\em \apj} { 279}, 763

\bibitem[\protect\astroncite{Press {\rm et~al.\/}}{1995}]{press}
Press, W., Flannery, B., Teukolsky, S., \& Vetterling, W., 1995,
\newblock {\em Numerical Recipies in C},
\newblock Cambridge University Press

\bibitem[\protect\astroncite{Randich}{1997}]{randich98}
Randich, S., 1997,
\newblock in R.~A. Donahue \& J.~A. Bookbinder (eds.), {\em Cool Stars, Stellar
  Systems and the Sun: Tenth Cambridge Workshop}, No. 154 in Astronomical
  Society of the Pacific Conference Series, p. 501, Astronomical Society of the
  Pacific

\bibitem[\protect\astroncite{{Reid} {\rm et~al.\/}}{1999}]{reid99dec}
{Reid}, I.~N., {Kirkpatrick}, J.~D., {Gizis}, J.~E., \& {Liebert}, J., 1999,
\newblock {\em \apjl} { 527}, L105

\bibitem[\protect\astroncite{{Spiegel} \& {Zahn}}{1992}]{spiegel92}
{Spiegel}, E.~A. \& {Zahn}, J.~., 1992,
\newblock {\em \aap} { 265}, 106

\bibitem[\protect\astroncite{{Tinney}}{1996}]{tinney96}
{Tinney}, C.~G., 1996,
\newblock {\em \mnras} { 281}, 644

\bibitem[\protect\astroncite{{Tinney}}{1998}]{tinney98discovery}
{Tinney}, C.~G., 1998,
\newblock {\em \mnras} { 296}, L42

\bibitem[\protect\astroncite{{Tinney} \& {Reid}}{1998}]{tinney98spec}
{Tinney}, C.~G. \& {Reid}, I.~N., 1998,
\newblock {\em \mnras} { 301}, 1031

\bibitem[\protect\astroncite{{Tinney} \& {Tolley}}{1999}]{tinney99}
{Tinney}, C.~G. \& {Tolley}, A.~J., 1999,
\newblock {\em \mnras} { 304}, 119

\bibitem[\protect\astroncite{Weiss}{1996}]{weiss94}
Weiss, N.~O., 1996,
\newblock in M.~R.~E. Proctor \& A.~D. Gilbert (eds.), {\em Lectures on Solar
  and Planetary Dynamos}, p.~59, Cambridge University Press, Cambridge

\end{thebibliography}

\begin{figure}[htb]
\caption{ \label{fig:sub4varicomp} Comparison between the 0.1-4.0 keV
X-ray light curve of \lp\ (top panel), and the 0.1-10.0 keV background
countrate (bottom panel).  Crosses mark the energies (right-hand scale)
of the detected counts.  The vertical broken lines mark the
(arbitrary) beginning and end of the flare; we refer the period prior
to this as the ``pre-flare'' period, and after this the ``post-flare''
period.  The variation observed in \lp\ is absent in the background
countrate taken from a much larger area on the same chip, but away
from the position of \lp. This demonstrates that the variability in
\lp\ is not due to variations in the background countrate in the chip.
}
\end{figure}

\clearpage
\pagestyle{empty}
\begin{figure}[htb]
\PSbox{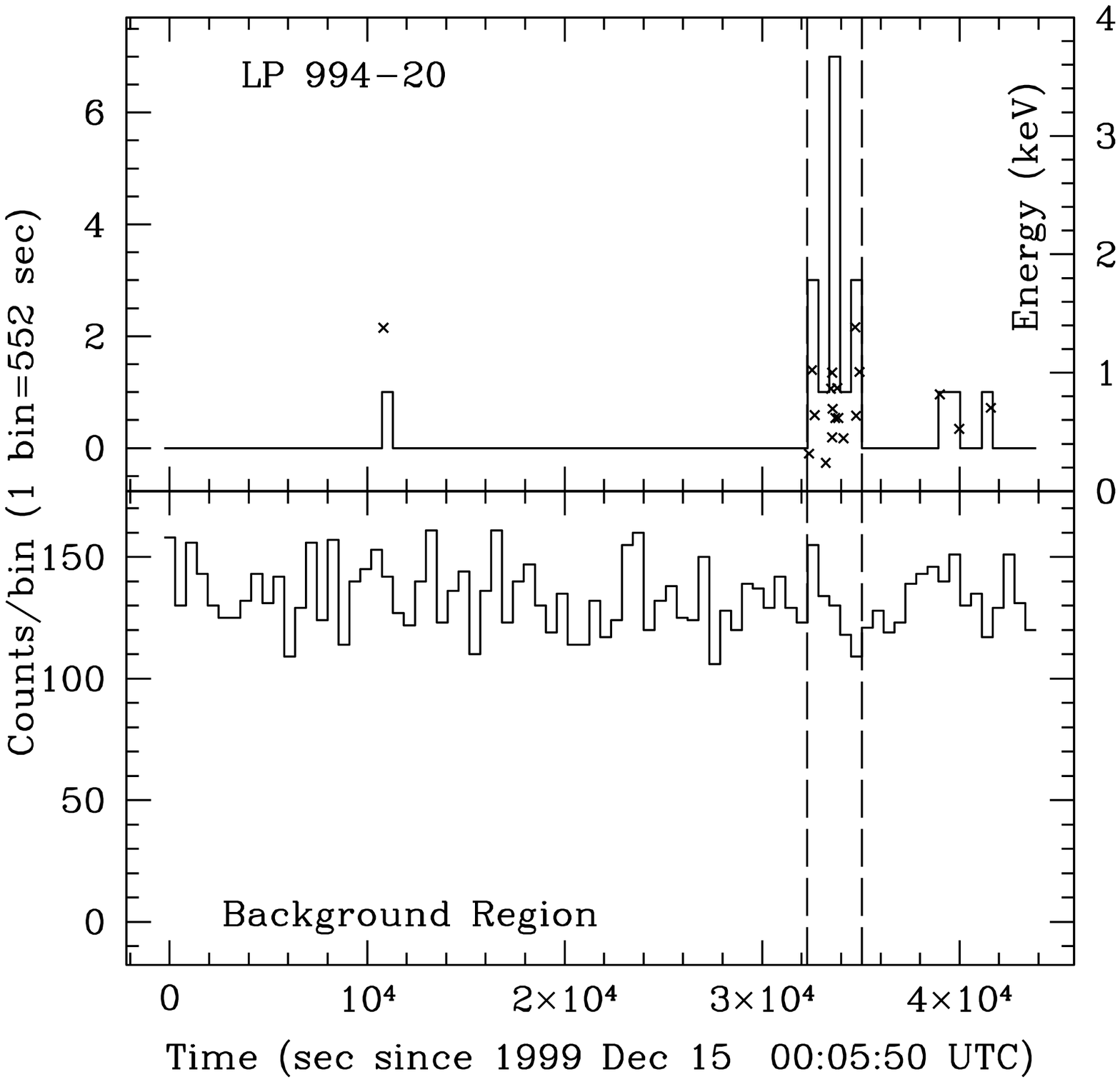 hoffset=-80 voffset=-80}{14.7cm}{21.5cm}
\FigNum{\ref{fig:sub4varicomp}}
\end{figure}

\begin{deluxetable}{lrr}
\scriptsize
\tablecaption{Derived Counts and \lxlbol\ \label{tab:flux}}
\tablewidth{7cm}
\tablehead{
\colhead{Period} & 
\colhead{\#Counts (bkg)}& 
\colhead{ $L_X/L_{\rm bol}$}}
\startdata
Full Obs.& 19 (0.57)	& 7\tee{-6}\\
Pre-flare& 1 (0.44)	& $<$2\tee{-6} \\
Flare	& 15 (0.14)	& 8\tee{-5} \\
Peak Flare& 7 (0.02)	& 2\tee{-4}\\
\enddata
\tablecomments{Assumed 1 count = 2.8\tee{-12}\erg \percm.   Upper limit
is 3$\sigma$. Source distance d=5.0 pc}
\end{deluxetable}

\end{document}